\documentclass[aps,12pt,showpacs]{revtex4}

\usepackage{graphicx}

\newcommand {\wa} {wavelet packet transform}
\newcommand {\bdm} {\begin{displaymath}}
\newcommand {\edm} {\end{displaymath}}

\begin{document}
\title{Adaptive multiresolution for wavelet analysis}
\author{R.~Sturani$^{1,2}$ and R.~Terenzi$^{3,4}$}
\affiliation{$^1$ D\'epartment de Physique Th\'eorique,
~Universit\'e~de~Gen\`eve,~Gen\`eve,~Switzerland\\ 
$^2$ INFN,~Presidenza~dell'INFN,~Roma,~Italy\\
$^3$ INFN,~Universit\`a~di~Roma ``Tor Vergata'',
Roma,~Italy\\
$^4$ INAF,~Istituto~Fisica~Spazio~Interplanetario,~Roma,~Italy\\
\texttt{e-mail: riccardo.sturani@physics.unige.ch,roberto.terenzi@cern.ch}}

\begin{abstract}
We present a new method of wavelet packet decomposition to be used in 
gravitational wave detection.
An issue in wavelet analysis is what is the time-frequency resolution which is
best suited to analyze data when in quest of a signal of unknown shape, like a
burst. In the other wavelet methods currently employed, like LIGO WaveBurst,
the analysis is performed at some trial resolutions.  
We propose a decomposition which automatically selects at any frequency the
{\em best} resolution. The criterion for resolution selection is based on 
minimization of a function of the data, named \emph{entropy} in analogy with 
the information theory.
As a qualitative application we show how a multiresolution time-frequency 
scalogram looks in the case of a sample signal injected over Gaussian noise. 
For a more quantitative application of the method we tested its efficiency as 
a non-linear filter of simulated data for burst searches, finding that it is
able to lower the false alarm rate of the WaveBurst algorithm with negligible 
effects on the efficiency.
\end{abstract}
\maketitle

\section{Introduction}
It is predicted that gravitational waves are emitted by a variety of 
cosmological and astrophysical sources and several detectors are now 
operating, or soon will, to directly observe a gravitational signal.
Among all possible kind of signals, bursts are generically
expected to be produced by compact astrophysical sources, like supernova 
explosions or compact object merging, as well as by high energy 
phenomena, like gamma rays bursts or more exotically by cosmic string cusps 
and kinks.
Not all of these processes are sufficiently well understood to predict a
specific gravitational waveform.
We use here ``burst'' generically, to designate a not better specified 
transient signals, with short duration, typically less than a second, and 
unknown shape.
The identification of bursts, possibly drowned in the detector noise, is
then an important goal in gravitational wave data analysis.
Solid detection algorithms are required to generate trigger of events out of
the time data stretch coming out of a detector.

Time-frequency decompositions for analysis of time series have been used 
since last decade in other research field than physics, see e.g. \cite{Canny}  
for an application on pattern recognition and \cite{Pol} for 
an application to physiological time series. More recently 
they have been applied to the gravitational wave detector data analysis, see 
e.g. \cite{Moh,Syl}. 
In particular the \emph{WaveBurst} algorithm \cite{Klim,Klim2}
has been employed in the the LIGO burst search, see also \cite{Ort} for
a wavelet method based on a different statistics than WaveBurst. 

Here we focus our attention on the issue of selection of the time-frequency 
resolution.
Since the signal one is looking for is unknown, it is a priori not clear 
what will be the resolution enabling to gather the power of the signal in the 
smallest possible number of wavelet coefficients, and consequently to better 
single it out from the noise. 
We then implemented a method to automatically recognize from the data 
themselves what are the best resolutions allowing to concentrate the power 
in the data in the least possible number of wavelet coefficients.
That is obtained by considering an \emph{entropy} function, defined so that 
it is maximum when the power is democratically distributed among all the
coefficients and minimum when all the power is concentrated in only one 
coefficient.

The plan of the paper is as follows. In Sec.~\ref{se:wdec} we explain
how we applied the adaptive multiresolution method, showing a case study
of a signal injected over Gaussian noise and how it is represented in the 
wavelet decomposition with varying time-frequency resolution.
In Sec.~\ref{se:roc} we use this method as a filter, showing that in 
combination with the WaveBurst algorithm it can be used to lower the false 
alarm rate, leaving almost unaffected the detection efficiency of the WaveBurst
trigger generator. We summarize our conclusion in Sec.~\ref{se:concl}.

\section{The wavelet decomposition and the entropy criterion}
\label{se:wdec}

We use a \wa\ to trade a set of data represented by a time series with
another set defined in the time-frequency domain. 

The output of a \wa\ can be represented by a binary tree where each node, 
labeled $\mathbf{W}_{i}^{j}$,  is an orthogonal vector subspace at
\emph{level} $j$ and at \emph{layer} $i$. 
Each of these levels will contain exactly the same information of the original 
time series, so that complete reconstruction of the original signal is
possible from any level. 

For instance, if the time resolution of the original series, made of $2^n$ 
samples, is $\Delta t_0$, information can be obtained from zero frequency up
to a maximal frequency
$f_{max}=(2\Delta t_0)^{-1}$ (we restrict here for simplicity to the case in 
which the number of sample is a power of two, but the method is completely 
general). 
Each level contains $2^n$ coefficients, arranged in $2^j$ layers, each of
which has $2^{n-j}$ coefficients. At level $j$ the time resolution is 
$\Delta t_j=2^{j}\Delta t_0$ and the frequency resolution is 
$\Delta f_j=2^{-j}f_{max}$. The coefficients in the $i^{\rm th}$ layer refer 
to the frequency bin characterized by $i2^{-j}<f/f_{max}<(i+1)2^{-j}$.

\begin{figure}[th]
\begin{center}
\includegraphics[width=.8\linewidth]{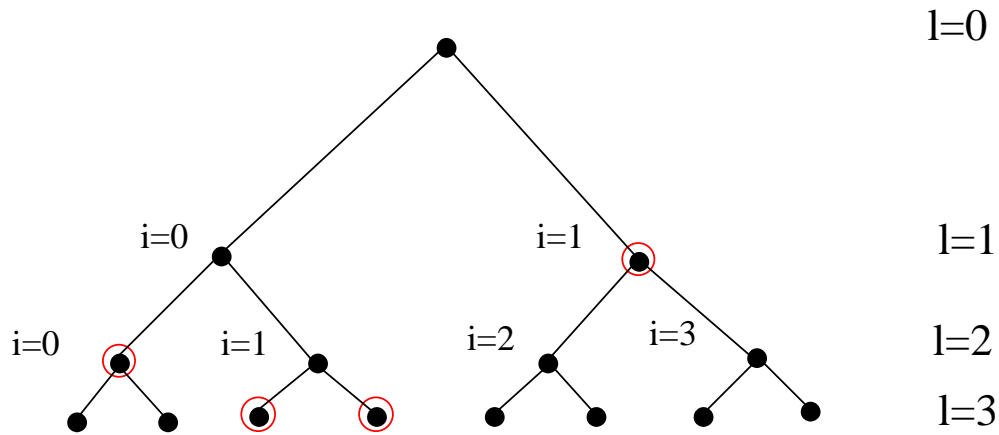}
\caption{Schematic example of a binary tree for a $l=3$ wavelet
  decomposition. The layers in each level are explicitly labeled for $l<3$. 
  The red circles show an example of a possible choice for wavelet packet
  basis. In this examples it is made by $W^2_0,W^3_2,W^3_3,W^1_1$.}
\label{fig:tree}
\end{center}
\end{figure}

Only a subset of these subspaces are needed to completely 
reconstruct the original time-signal: it can be shown (see e.g. \cite{Mal})
that the set of \emph{leaves} of every {\it admissible tree} completely 
represents 
the original signal in the wavelets domain, where an admissible tree is a 
sub-tree of the original binary decomposition tree where every node has 
exactly $0$ or $2$ children nodes (see e.g. fig.~\ref{fig:tree}, where the
leaves of an admissible tree are marked by red circles).\\
Having such a redundancy, how to choose the set of  $\mathbf{W}_{i}^{j}$ to 
represent the signal? In \cite{Klim} the $\mathbf{W}_{i}^{j}$ are chosen 
setting $j=L$, where $L$ is the {\it decomposition level} and consequently 
$i=0,\ldots,2^{L-1}$. Here we present a way to choose the 
$\mathbf{W}_{i}^{j}$ based on an {\it entropy } function $E^{2}(W_{i}^{j})$, 
where $ W_{i}^{j}$ is the set of the coefficients of the base 
$\mathbf{W}_{i}^{j}$ of a node in the decomposition binary tree 
$\mathcal{T}$, and defined as follows
\begin {equation}
E^{2}(W_{i}^{j})= -\sum_{k} {x^{2}_{k} \over{||X||^{2}}} \log_{2} {x^{2}_{k} 
\over{||X||^{2}}}\,,
\label{eq:entropy}
\end{equation}
where the $x_{k}\in W_{i}^{j}$ are the \wa\ coefficients of a 
$\mathbf{W}_{i}^{j}$ node and 
\begin {equation}
||X||^{2}= \sum_{i} {x_{i}^{2}}\,.
\end{equation}

From  the complete decomposition binary tree $\mathcal{T}$ of a \wa, we select 
the admissible tree $\mathcal{A}_{E}\subset\mathcal{T}$ whose leaves 
$\mathbf{W}_{i}^{j}\in\mathcal{A}$ have the minimum cost, i.e. where 
$E^{2}(W_{i}^{j})$ is minimum for all the children nodes of $W_{i}^{j}$ in the 
original \wa\ binary tree $\mathcal{T}$ ({\it entropy criterion} \cite{Wic}, 
where a different notation is used).

From among all the admissible binary trees, $\mathcal{A}_{E}$ is the one that 
represents the signal most efficiently. By ``efficient" we mean that a signal 
can be represented by a small number of  wavelet packets, that is, the basis
for the decomposition is chosen such that the weight of the coefficients is 
concentrated on a small number of wavelet packets and a large number of 
coefficients are close to zero.
In fact from eq.(\ref{eq:entropy}) we can see that  if in a particular 
basis the decomposition produces all zero coefficients except one 
(i.e., the signal coincides with a \wa\ wave form), then the entropy reaches 
its minimum value of zero. On the other hand, if in some basis the 
decomposition coefficients are all equally important, say $x_{k} = 1/N$ where 
$N$ is the length of the data, the entropy in this case is
maximum, $\log_{2}N$.
Any other decomposition will fall in between these two extreme cases. In 
general, the smaller the entropy the fewer significant coefficients are needed 
to represent the signal.

\begin{figure}[th]
\begin{center}
\includegraphics[width=.8\linewidth]{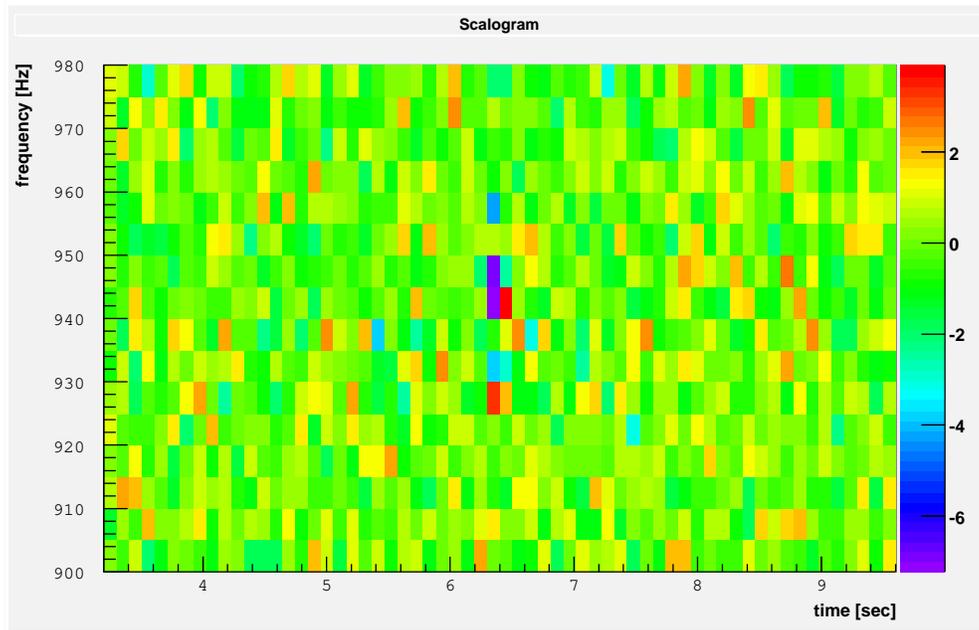}
\caption{Scalogram in the time-frequency plane. The diagram shows white noise
  data with an injection $h(t)=A e^{-t^2/(2\Delta^2)}\sin(2\pi f)$, 
  with $t_0=6.4$sec, $\Delta=0.06$sec, $f=930$Hz and $A\simeq0.76$. The 
  wavelet transform is made on a Symlet basis with maximum decomposition level 
  equal to 9.}
\label{fig:scalogram}
\end{center}
\end{figure}

\begin{figure}[th]
\begin{center}
\includegraphics[width=.8\linewidth]{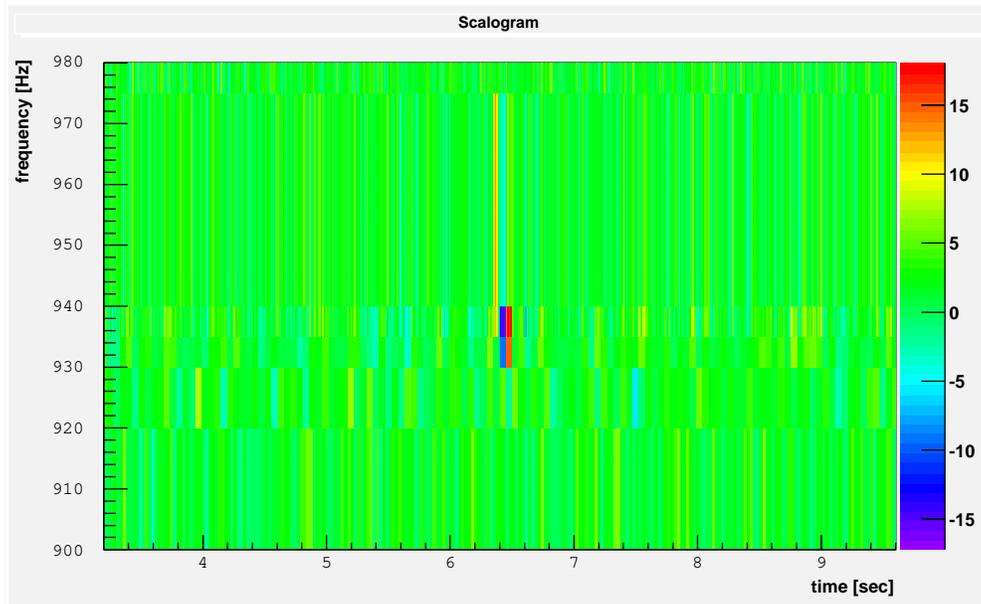}
\caption{Scalogram in the time-frequency plane as in fig.~\ref{fig:scalogram}, 
but from entropy multiresolution method.}
\label{fig:scalogramM}
\end{center}
\end{figure}

As an illustration we show in figs.~\ref{fig:scalogram} and 
\ref{fig:scalogramM} two time-frequency decompositions, or \emph{scalograms}. 
In fig.~\ref{fig:scalogramM} we report an example of multiresolution
scalogram, where we consider a time series with a sampling time of $0.2$msec,
and then performed a wavelet transform up to level 9 using our adaptive
multiresolution algorithm which automatically detects the resolution which 
is better capable to ``concentrate'' the signal. 
As a comparison in fig.~\ref{fig:scalogram} we show the same data in a 
time-frequency analysis done at decomposition level 9.

\section{Test of the detection algorithm}
\label{se:roc}
Within the context of a \wa\ several detection procedures has been developed 
(see e.g. \cite{Klim,Klim2} and \cite{Ort}). 
Here we adopted the WaveBurst method as in \cite{Klim} and we show a possible 
application of the entropy criterion to improve the method.
The WaveBurst algorithm is a trigger event generator, it identifies
candidate events by looking for power excess in a wavelet decomposition and
collecting clusters of higher coefficients.
We suggest, as a possible application of our entropy multiresolution method, 
that a signal can be filtered by looking for excess power not in the wavelet
decomposition with some given resolution (i.e. by choosing a basis belonging
to the same decomposition level), but in the wavelet decomposition 
which makes use of the entropy method, see fig.~\ref{fig:tree}.
After having zeroed the coefficients under a given threshold, one can 
reconstruct from the surviving wavelet coefficients a new, filtered, time 
series which can be fed to the WaveBurst algorithm.

The original and the filtered data are both time series and the WaveBurst 
algorithm can be run on both of them. We expect that if a signal is present in
the original series, it should be stable under this analysis, or better, that 
the entropy-multiresolution method can catch all of the signal with a
different noise distribution over the wavelets coefficients. So we can check
the coincidences between the event found by WaveBurst on the original time
series and on the entropy-filtered one.

We simulated one hour of white noise data, to which we added 71 injections
separated by 50 seconds. The injections had the temporal profile
\bdm
h(t)=A\exp\left(-\frac{t^2}{2\Delta t^2}\right)\sin(2\pi f t)\,,
\edm
with $f=930$Hz, $\Delta t=0.06$ sec and $A\simeq 0.76$.
\vspace{1.1cm}
\begin{figure}[th]
\begin{center}
\includegraphics[width=.8\linewidth]{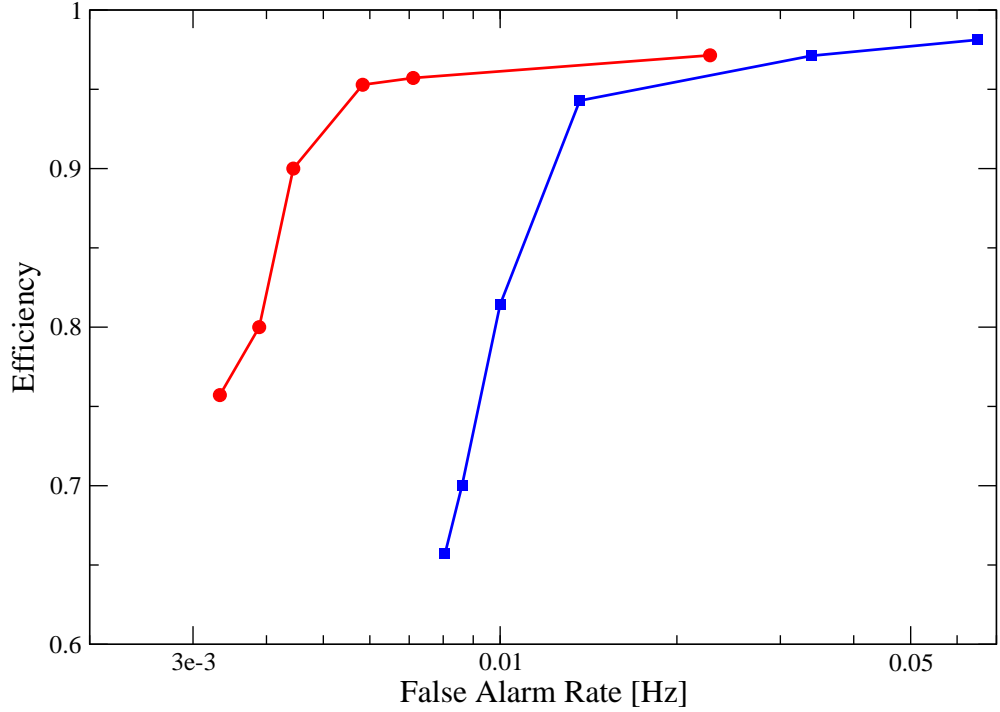}
\caption{ROC curves for WaveBurst method applied on a time stretch of white 
data with sine Gaussian injections. The blue squares refer to the straight
WaveBurst algorithm, the red circles to the coincidences between the events 
found by WaveBurst on the original data and on the data filtered after the 
adaptive multiresolution decomposition. 
On the ROC curves the percentage of coefficients which are zeroed varies.}
\label{fig:roc}
\end{center}
\end{figure}

We first considered how many injections were recovered (efficiency) and how 
many false alarms were found by applying the WaveBurst method to the original 
data to draw a Receiver Operating Characteristic (ROC).
We then considered the above mentioned coincidences between the events found
on the original data and the events found on the data which had undergone
our adaptive entropy multiresolution filter. The two ROC curve are shown in 
fig.~\ref{fig:roc} and they show how the coincidence method enables to lower
substantially the false alarm rate without affecting the efficiency.

\section{Conclusions}
\label{se:concl}

We presented a new method of wavelet packet decomposition to be used in 
gravitational wave detection.
This entropy multiresolution method is aimed to better represent in the 
wavelets domain a signal embedded in noise in order to built a better filter 
in this domain.
We presented also one of the possible applications of this method, showing how 
to use it in order to have an improved ROC curve by well tested methods, 
like WaveBurst.
In following work we hope to present other applications of this entropy 
multiresolution method.

\section*{Acknowledgments}
The authors wish to thank the ROG collaboration and A. Ortolan 
for useful discussions. R.T. wishes to thank G.V. Pallottino for his
encouraging conversations during the study and development of the 
entropy-multiresolution method and R.S. wishes to thank G. Vedovato 
and M. Drago for their invaluable help and stimulating discussions.

\end{document}